\documentclass{article}

\usepackage[preprint,nonatbib]{nips_2018}

\usepackage[utf8]{inputenc} % allow utf-8 input
\usepackage[T1]{fontenc}    % use 8-bit T1 fonts
\usepackage{hyperref}       % hyperlinks
\usepackage{url}            % simple URL typesetting
\usepackage{booktabs}       % professional-quality tables
\usepackage{amsfonts}       % blackboard math symbols
\usepackage{nicefrac}       % compact symbols for 1/2, etc.
\usepackage{microtype}      % microtypography
\usepackage{textcomp}
\usepackage[symbol]{footmisc}
\usepackage{cleveref}

\usepackage{subcaption}

\usepackage{graphicx}

\usepackage{booktabs}
\usepackage{color}

\newcommand{\secref}[1]{Section~\ref{#1}}

\newcommand{\figref}[1]{Figure~\ref{#1}}
\newcommand{\eqref}[1]{Equation~\ref{#1}}

\newif\ifcomments
\commentsfalse
\ifcomments
\newcommand{\alexey}[1]{\textcolor{blue}{AT: #1}}
\else
\newcommand{\alexey}[1]{\textcolor{blue}{}}
\fi

\ifcomments
\newcommand{\paras}[1]{\textcolor{blue}{PJ: #1}}
\else
\newcommand{\paras}[1]{\textcolor{blue}{}}
\fi

\ifcomments
\newcommand{\ajay}[1]{\textcolor{blue}{AJ: #1}}
\else
\newcommand{\ajay}[1]{\textcolor{blue}{}}
\fi

\title{Dynamic Space-Time Scheduling for GPU Inference}
\author{
Paras Jain, Xiangxi Mo, Ajay Jain$^{\S}$,\\
\textbf{Harikaran Subbaraj, Rehan Sohail Durrani,} \\
\textbf{Alexey Tumanov, \textbf{Joseph Gonzalez}, \textbf{Ion Stoica}} \\
  University of California, Berkeley \\
  $^\S$Massachusetts Institute of Technology \\
\texttt{\{paras\_jain, xmo, hsubbaraj, rdurrani, atumanov,} \\ \texttt{jegonzal, istoica\}@berkeley.edu},   \texttt{ajayjain@mit.edu}
}

\usepackage[dvipsnames]{xcolor}

\newcommand{\todo}[1]{}
\renewcommand{\todo}[1]{{\color{BrickRed} [ TODO: {#1} ] }}

\begin{document}

\maketitle
%\footnotetext{$^*$ Authors listed in alphabetical order}

\begin{abstract}
% \todo{Paras rewrite} 
Serving deep neural networks in latency critical interactive settings often requires GPU acceleration.
However, the small batch sizes typical in online inference results in poor GPU utilization, a potential performance gap which GPU resource sharing can address.
In this paper, we explore several techniques to leverage both temporal and spatial multiplexing to improve GPU utilization for deep learning inference workloads. 
We evaluate the performance trade-offs of each approach with respect to resource-efficiency, latency predictability, and isolation when compared with conventional batched inference.
Our experimental analysis suggests up to a 5x potential for improved utilization through the exploration of more advanced spatial and temporal multiplexing strategies. Our preliminary prototype of a dynamic space-time scheduler demonstrates a 3.23x floating-point throughput increase over space-only multiplexing and a 7.73x increase over time-only multiplexing for convolutions, while also providing better isolation and latency predictability.
\end{abstract}

% The abstract paragraph should be indented \nicefrac{1}{2}~inch
% (3~picas) on both the left- and right-hand margins. Use 10~point
% type, with a vertical spacing (leading) of 11~points.  The word
% \textbf{Abstract} must be centered, bold, and in point size 12. Two
% line spaces precede the abstract. The abstract must be limited to
% one paragraph.
\section{Introduction}
\label{sec:intro}

% Deep learning today is an ubiquitous machine learning technique with state-of-the-art performance on common computer vision tasks such as image classification \citeme{} and in natural language processing applications \citeme{}. 
GPUs are essential to deep learning.
By leveraging substantial parallelism, high memory bandwidth, and tensor acceleration, GPUs are able to quickly compute activations over large batches of inputs.
NVIDIA’s datacenter-class V100 GPU, for example, packs more than 120 TFLOP/s of half-precision matrix multiply-and-accumulate performance designed specifically for deep learning workloads.
% Their ability to quickly compute gradients in parallel over large batches of input have made them 
As deep learning is deployed in applications ranging from video monitoring, language translation, and speech recognition, there is an emerging need for parallel hardware accelerators to support inference.
While there are numerous specialized inference processors, the widespread availability of GPUs and their support for general deep learning models renders GPUs indispensable for inference.

\begin{figure}[!tbp]
  \begin{minipage}[b]{0.5525\textwidth}
    \centering
    \includegraphics[width=\textwidth]{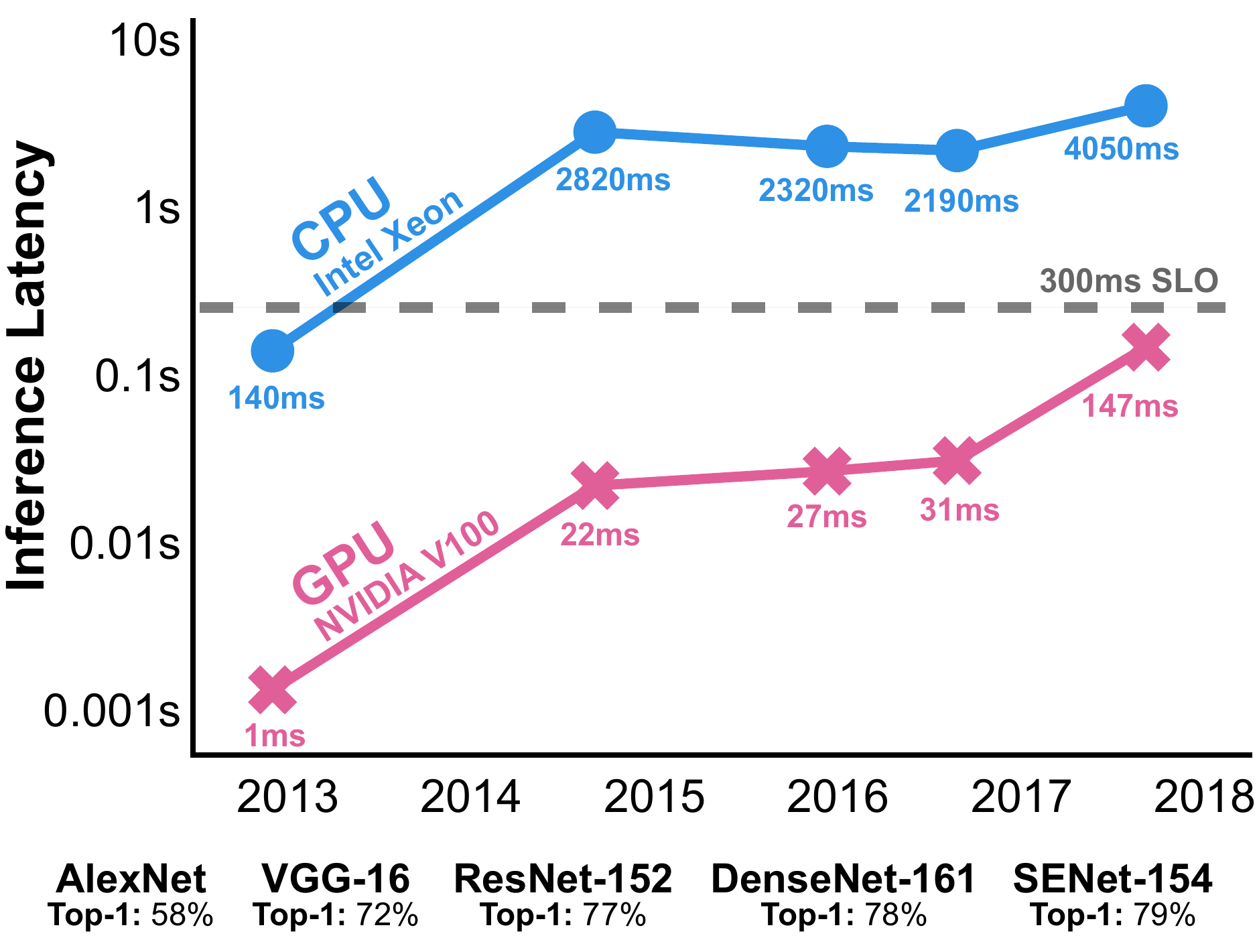}
    \caption{DNN model complexity is increasing over time on CPUs and GPUs. Most models fail to meet the 300ms latency SLO on a CPU. Models from left-to-right: AlexNet \cite{krizhevsky2012imagenet}, VGG-16 \cite{Simonyan14c}, ResNet-152 \cite{he2016deep}, DenseNet-161 \cite{huang2017densely} and SENet-184 \cite{hu2018senet}.}
    \label{fig:intro:latency}
  \end{minipage}
  \hspace{0.025\textwidth}
  \begin{minipage}[b]{0.4125\textwidth}
    \centering
    \includegraphics[width=\textwidth]{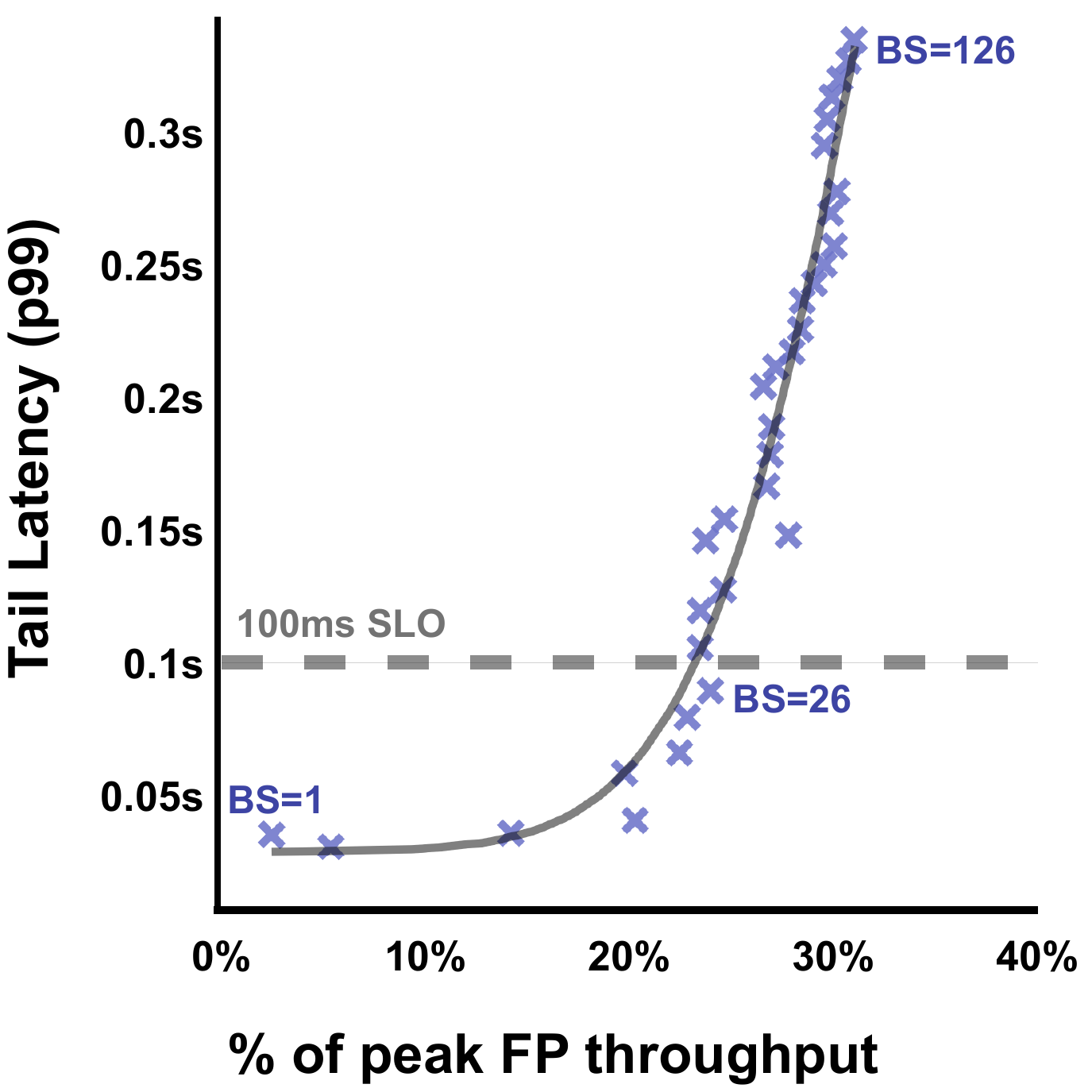}
    \caption{In order to meet latency SLOs, small batch sizes must be used resulting in low GPU utilization. The largest batch size for ResNet-50 (plotted) within the SLO is 26, but only achieves an average of 28\% of peak V100 FP32 throughput.}
    \label{fig:intro:batchsize}
  \end{minipage}
\end{figure}

DNN training is computationally expensive but relatively infrequent, while online inference needs to scale to billions of queries per day and is rapidly outpacing training in datacenters~\cite{hazelwood2018applied}.
Amazon recently announced~\cite{reinvent} that roughly 90\% of the machine learning computation is spent on inference (not training).
% \joey{Can someone verify what the 90\% measures all computation or machine learning and is this in terms of cpu hours, money.}
Key metrics for revenue-critical applications can dramatically suffer with an increased application latency~\cite{schurman_brutlag}.
In spite of tight end-to-end latency budgets ($<100\textrm{ms}$), we note an alarming trend that inference latency on a CPU has been on a rise~(\figref{fig:intro:latency}). The researcher's appetite for better accuracy leads to ever-increasing model size and complexity; as an example, the state-of-the-art SENet-184~\cite{hu2018senet} model has a $4.1\textrm{s}$ CPU inference latency.
Given that increases in interactive query latencies leads to losses in revenue, CPUs cannot support today's interactive model serving workloads which leaves GPUs an obvious favorite.

While training workloads continue to scale and can often easily saturate modern GPUs, ML inference has distinctly different performance requirements that often result in poor GPU utilization. In contrast to throughput-oriented model training, revenue-critical inference workloads must meet latency objectives with queries often arriving stochastically.
In practice, online inference queries often cannot realize the high levels of parallelism that offline iterative minibatch training achieves; lower parallelism leads to poor GPU utilization in practice.

Small batch sizes are an unfortunate reality for online inference with SLOs~(\figref{fig:intro:batchsize}), which leads to low utilization.
% Small batches are common in online inference as queries must be dispatched without waiting to buffer a batch of requests due to strict latency objectives.
The issue raised by small inference batch sizes is exacerbated as model complexity grows over time, pushing GPU inference latencies to approach interactive SLOs from below (as noted in \figref{fig:intro:latency}).
Given that inference workloads must run continuously and respond to highly variable demand, capacity must be provisioned for demand peaks which lowers GPU utilization even further.

The current practice of exclusive access to a GPU cannot scale due to the low utilization on current hardware. We notice that small batch sizes can lead the GPU to low utilization under 15\%. This problem affects other throughput-oriented accelerators; Google's Tensor Processing Unit reports an observed throughput under 23\% of the peak on average and under 4\% for some models~\cite{tpuISCA17}.

A common approach to improving utilization of parallel hardware, under stochastic query load, is to leverage multi-tenancy.
By sharing a GPU across multiple prediction workloads we can potentially leverage workload level parallelism and achieve statistical multiplexing.
However, leveraging multi-tenancy on a GPU remains an open research problem. 
First, its runtime performance must be \textit{predictable}, which, as we show in \secref{sec:eval}, is not always true. 
Second, it must be \textit{resource-efficient}. To address this, we explore a number of techniques for sharing a GPU among a set of execution kernels, each with their drawbacks.
Third, a measure of \textit{performance-isolation} is needed, which is typically achieved through fair resource allocation.

Current approaches for sharing a GPU for DNN inference either multiplex the GPU across space or across time.
\secref{sec:space_time_multiplex_limits}~evaluates current approaches against the three criteria established above. We argue that only multiplexing across space or time leads to a compromise on one of these criteria -- instead, we propose scheduling across space and time for GPU inference in~\secref{sec:proposed}.
By packing multiple execution kernels across disjoint DNN graphs with dynamic query batching, we show potential for a multi-tenancy solution that is resource-efficient (\(>\) 3x increase in throughput over state-of-the-art) while providing isolation and predictability.

%\paras{write space-time story}
%%%
%This calls for techniques to increase GPU utilization. Obvious state-of-the-art choices:
%* space multiplexing
%* time multiplexing 
%Multiplexing the GPU along any of the two dimensions is insufficient. In this paper, we show that and call for space-time scheduling for GPU inference.
%
%\alexey{cite TPU paper again with low utilization (10\%) which implies 10x opportunity gap} \paras{Take this sentence out, it spoils the punchline} As a consequence, low device utilization per inference request with expensive models subject to tight latency constraints calls for a solution that involves multiplexing the GPUs across several simultaneously executing models (space) and over continuously streaming workloads (time).
%%%

%\paras{Key contributions?} In this work we explore several techniques to leverage multi-tenancy in the context of inference. (1) \textit{Time Multiplexing} -- multiplexes a GPU by timeslicing it between a specified set of execution kernels. (2) \textit{Spatial Multiplexing} (CUDA Streams) -- uses multiple CUDA streams \cite{nvidia_hyperq} for concurrent execution of several kernels together. And (3) \textit{DNN-aware kernel scheduling} -- our proposed approach. In light of the shortcomings of existing GPU multiplexing mechanisms, we make a case for packing multiple execution kernels, with dynamic query batching, as we show that spatial-only or temporal-only GPU multiplexing is insufficient, calling for a spatio-temporal GPU kernel scheduler.

\section{Application Model: A Managed Cloud Inference Service}
\label{sec:systemmodel}

Consider a cloud-based managed service for deploying ML models for online inference, similar to Amazon AWS SageMaker~\cite{web_sagemaker} or Google Cloud ML Engine~\cite{google_mlengine}. Users may develop and upload their trained machine learning models to this service. The service then deploys the model onto one-or-more replicas, which each may use CPUs and GPUs. The service and the users agree on some Service Level Objective (SLO), such as a measure of tail-latency for model inference.

We simplify this model in order to isolate interference effects due to multi-tenant execution. First, all models running on a single GPU are restricted to the same architecture (but different weights). 
This separates the impact of heterogeneous model architectures from multi-tenancy. 
Second, request queues are always saturated, thereby isolating model service latency from request queuing latency.
It is worth noting that addressing both model heterogeneity and queuing latency is a key focus of future work.

\section{Investigating limitations of space-only and time-only multiplexing}
\label{sec:space_time_multiplex_limits}

Using this simplified model of a managed cloud inference service, we outline three leading approaches to model inference today:

% \begin{enumerate}
% \item 
\textbf{Exclusive access.} \label{V1} Each model has an exclusive GPU. Amazon AWS SageMaker, Google Cloud ML Engine, Clipper~\cite{crankshaw2017clipper} and TensorFlow Serving~\cite{tf-serving} all utilize this approach. In this approach, inference is done in batches. When the network is performing forward propagation, new queries must wait in a queue until one forward pass is completed. 

% \item 
\textbf{Time Multiplexing.} An on-device scheduler enables interleaved execution of multiple CUDA contexts at once. This approach is common when multiple processes run concurrently using the same GPU. This approach relies on the kernel to time-multiplex processes and GPU to swap contexts when different processes compete for the same resource. 

% \item 
\textbf{Spatial Multiplexing.} Kernel execution can overlap by utilizing NVIDIA Hyper-Q~\cite{nvidia_hyperq}. CUDA Streams and NVIDIA Multi Process Service (MPS)~\cite{nvidiamps} utilize multiple hardware queues to enable spatial sharing of the GPU. 
The CUDA Streams API is used by ModelBatch~\cite{narayanan2018modelbatch} and NVIDIA TensorRT~\cite{nvidia_tensorrt}. AMD's MxGPU (SR-IOV)~\cite{amdmxgpu} is another approach for spatial multiplexing, not considered in this work. In this paper, we consider two kinds of spatial multiplexing:
\begin{enumerate}
    \item \textit{Implicit Spatial Multiplexing with MPS}: NVIDIA MPS allows multiple processes to run on the device at the same time by allocating them different cuda streams. 
    \item \textit{Explicit Spatial Multiplexing with CUDA Streams}: With this method we directly interact with multiple CUDA streams inside a single processes. 
\end{enumerate}

% \end{enumerate}

We examine competitive solutions for each of these model inference approaches to identify if they satisfy our chosen system criteria, namely: \textit{latency predictability}, \textit{resource-efficiency}, and \textit{performance isolation}.

\subsection{Experimental Setup}
\label{subsec:experimentalsetup}
We evaluate these three virtualization methods on two image classification neural network architectures: MobileNet V2~\cite{sandler2018mobilenetv2} and ResNet-50~\cite{res50}. These two models are popular choices for low-compute and high-accuracy classification applications respectively.

\begin{enumerate}
    \item \textit{Exclusive access} is tested with a single model executing batched queries on a private GPU. Although we cannot use this approach with multiple models on a single GPU, this test represents a single-tenant lower bound on latency and an ideal best-case for performance.
    \item  \textit{Time multiplexing} is tested by running each model in a separate CUDA context and utilizing a software scheduler to interleave execution. This provides memory safety and some basic level of isolation between tenants.
    \item \textit{Spatial multiplexing} is tested by using the NVIDIA MPS server to partition model queries for different models across a pool of CUDA streams.
\end{enumerate}

All experiments use p3.2xlarge or p3.8xlarge instances on Amazon AWS. These instances have direct access to NVIDIA V100 datacenter-class GPUs with up to 14 TFLOP/s of single-precision floating-point throughput.
%\ajay{p3.2xlarge has no NVLink (14 TFLOP/s). p3.8xlarge has NVLink with 15.7 TFLOP/s throughput}
We do not test Tensor Core sharing in our experiments.
\subsection{Preliminary results}
\label{sec:eval}

\begin{figure}[!t]
    \centering
    \includegraphics[width=\textwidth]{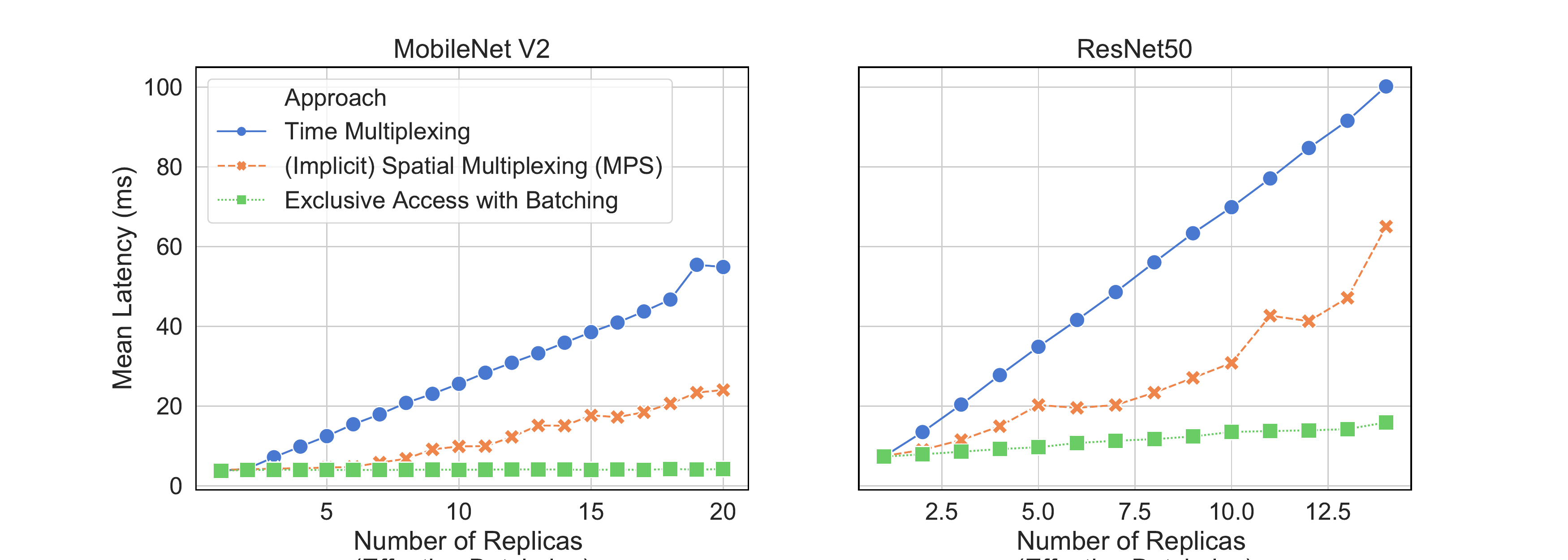}
    \caption{\textbf{Both time and spatial multiplexing do not meet the performance of exclusive access; however, spatial multiplexing is able to deliver better inference latency than time multiplexing.}  We compare three approaches to GPU multi-tenancy. Exclusive access (modeled by batching a single model) provides fast and predictable latencies at a high cost. Time multiplexing dramatically increases inference latency as sharing increases. Spatial multiplexing through NVIDIA MPS better manages latency by sharing resources.}
    % \ajay{Why is mean latency plotted, not P99?}
    \label{fig:eval:latency}
\end{figure}

\begin{figure}[!b]
  \begin{minipage}[b]{0.49\textwidth}
    \centering
    \includegraphics[width=\textwidth]{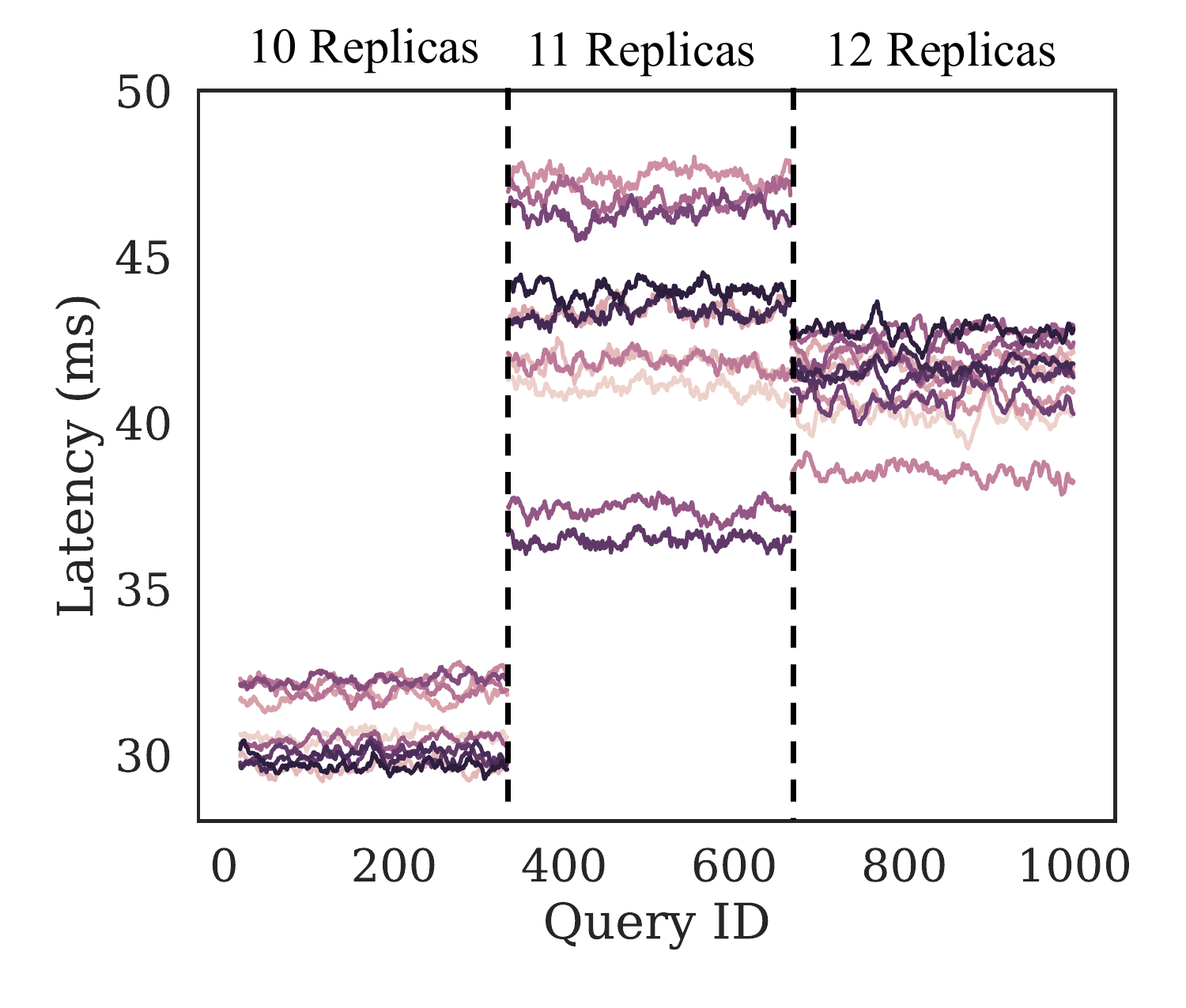}
    \caption{\textbf{Implicit spatial multiplexing (with MPS) has unpredictable latency when different number of processes are running concurrently.} As we add replicas to a GPU running 10 multi-tenant models, we observe unpredictability, which we suspect is caused by the on-device scheduler.}
    \label{fig:eval:unfair}
  \end{minipage}
  \hspace{0.025\textwidth}
  \begin{minipage}[b]{0.48\textwidth}
    \centering
    \includegraphics[width=\textwidth, height=6cm]{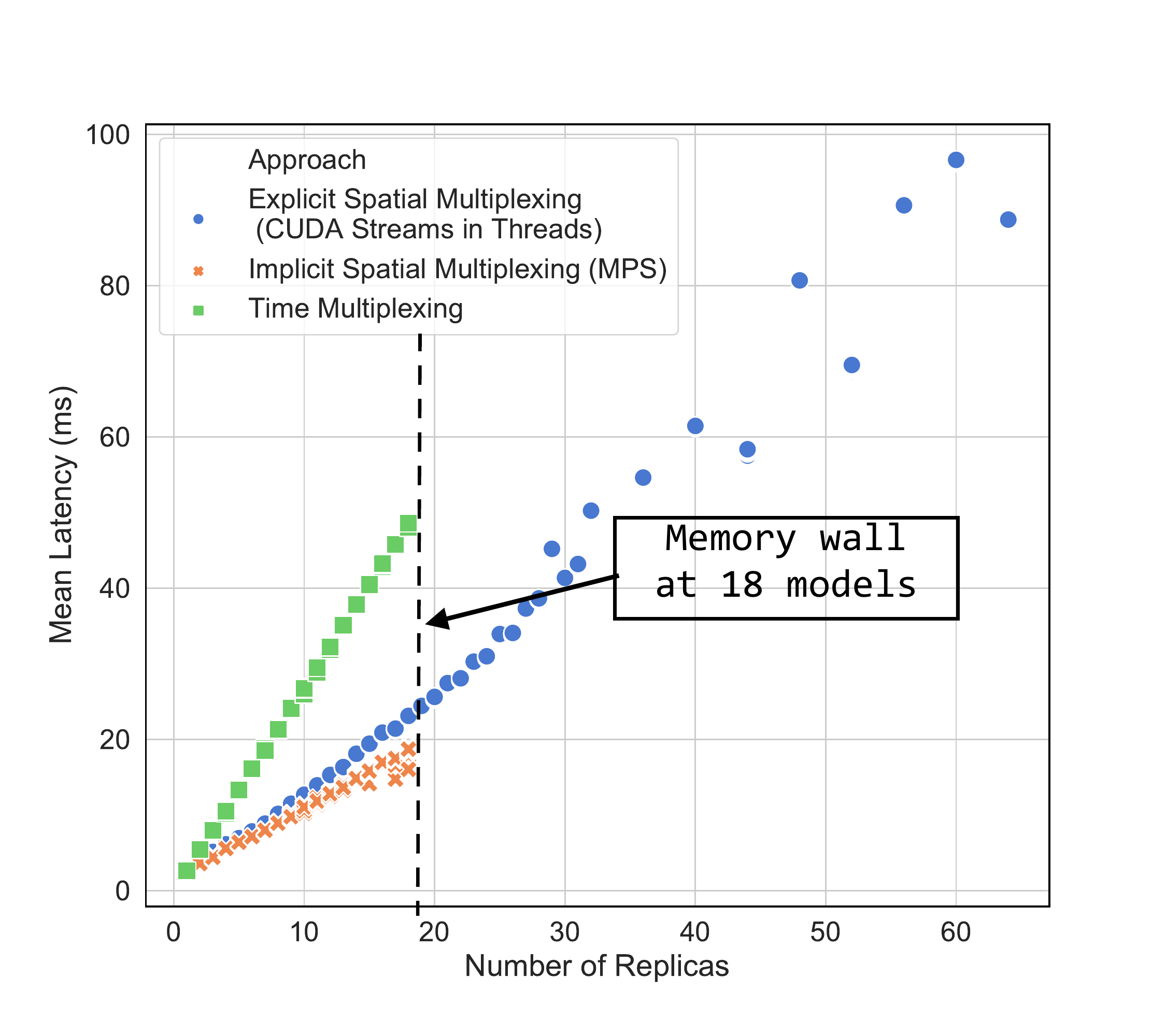}
    \caption{\textbf{Both time and implicit spatial multiplexing are bounded by memory. Explicit spatial multiplexing is not.} In this experiment, most approaches hit a 16 GB memory wall at 18 replicas, at which point GPU memory was exhausted; however, explicit spatial multiplexing (CUDA Streams on different threads), was able to scale up to at least 60 ResNet-50 models.}
    \label{fig:eval:memory}
  \end{minipage}
\end{figure}

We report results from our benchmark in~\figref{fig:eval:latency}. For both MobileNet V2 (compute-optimized model) and ResNet-50 (high-accuracy model), we tested batched exclusive access, time multiplexing, and spatial multiplexing. Batched exclusive access devotes the entire GPU to a single model, unlike the other two approaches. This is an extremely aggressive baseline and represents the ideal performance a model would achieve if it were the only tenant and throughput is our only objective.  However, if we also wanted to minimize latency we would likely use much smaller batch sizes. 

%Time-only multiplexing suffers a \todo{xx} slowdown compared to exclusive access at 20 replicas 
Overall, time-only multiplexing suffers a geometric mean 4.6x slowdown compared to exclusive access 
% \joey do you mean "to the projected fractional performance"
% (i.e. $L_{\small\textrm{exclusive}} / L_{\small\textrm{time}}$)
while space-only multiplexing only endures a 2.2x slowdown, across the experiments in \figref{fig:eval:latency}.
Ultimately, no single solution wins on all criteria we established in \secref{sec:intro}.

\textit{Exclusive access}, as discussed, allows for high-throughput, low-latency, isolation, and predictability, but is extremely expensive, since it does not share the GPU at all. As we note in \figref{fig:intro:batchsize}, this increase in performance comes at a tradeoff - namely that online inference workloads must endure low GPU utilization in order to meet these tight latency SLOs. We believe this is a great model for users who have high enough request rates during inference to achieve good utilization on a GPU and are able to batch requests to arrive simultaneously into one forward pass of the network.
%\ajay{Isn't this recommending a batch approach? The exclusive access model is low throughput in the batch-size 1 case}

\textit{Time multiplexing} (CUDA context switching) can actually accomplish multi-tenancy with good isolation and predictability, but at the cost of degraded throughput and high latencies. The main drawback to this approach is that it cannot take advantage of parallel execution of the kernels, since the GPU only allows one running CUDA context at a time. This approach instead interleaves processes resulting in slightly improved resource-efficiency; although it still suffers from poor utilization during each schedule quantum, explaining the linear-slowdown as the number of replicas grows. Poor latency scalability makes time multiplexing alone an inadequate solution for interactive inference query serving.

\textit{Spatial multiplexing} (Hyper-Q) does improve on poor utilization and achieves much better resource-efficiency. However, we find there is poor predictability and isolation in this setup. It seems the spatial multiplexing approach is extremely sensitive to the choice of the number of tenants. Each tenant appears to have fairly consistent behaviour once the model runs. But, across multiple model tenants, there is up to a 25\% latency gap between the fastest model on a GPU and the slowest straggler model as seen in \figref{fig:eval:unfair}. Furthermore, the unpredictability and discrepancy between latencies across different processes is exacerbated when an odd number of processes runs concurrently with MPS enabled.
\section{A new hope? Dynamic space-time scheduling}
\label{sec:proposed}

From our evaluation of current approaches, we find that neither time-only multiplexing nor space-only multiplexing can meet all three performance criteria: good resource efficiency, isolation and predictability. GPU single-tenancy leads to poor utilization and high costs~(\figref{fig:intro:batchsize}). \figref{fig:eval:latency} and \figref{fig:eval:memory} demonstrate inherent scalability limitations to common space-only and time-only multiplexing strategies. \figref{fig:eval:unfair} details unpredictable latency as tenants are added to a GPU. \figref{fig:eval:memory} shows  the only way to schedule hundreds of models on GPU is to use single process utilizing one CUDA stream per thread; since we are micro-managing the inference by dispatching kernels to different streams, there is an opportunity for more fine-grained scheduling control over the streams to optimize latency and throughput. 

In light of these limitations, we propose a promising new approach we call \textit{dynamic space-time scheduling}. 
The key idea is trade-off space and time multiplexing in order to efficiently utilize the GPU while preserving isolation and predictability.

We preserve predictability and isolation during virtualization by monitoring inference latencies per-kernel. This allows reallocating resources between tenants on-the-fly. Moreover, we notice that CUDA Stream scheduling anomalies typically only create a few stragglers, so we can simply evict degraded workers without significantly impacting total system throughput. We are further investigating this approach in ongoing work.

Our approach also dramatically improves resource-efficiency on the GPU -- we observe a 7.71 overall geometric mean speedup in throughput compared to time-only multiplexing and a 3.23 speedup compared to space-only multiplexing, as shown in~\figref{fig:proposed:sgemm}. Space-time scheduling merges many concurrent small kernels from  disjoint DNN graphs into a small set of larger super-kernels that together fill the GPU. The super-kernel avoids the scheduling penalty associated with current space-only multiplexing approaches.

As interactive inference queries arrive stochastically, we cannot easily precompute super-kernels ahead-of-time. Instead, the space-time scheduler must dynamically schedule kernels as they arrive. We are investigating the design of a more general dynamic scheduler, but we notice that overheads gradually decrease if we cache super-kernels as workloads stabilize over time.

Our goal is to optimize a batch of distinct models dynamically, in comparison to ahead-of-time DNN graph optimizers like TVM~\cite{tvm}, Tensor Comprehensions~\cite{tensor-comprehension}, Halide~\cite{halide}, GLOW~\cite{GLOW} and TensorRT~\cite{nvidia_tensorrt}. These optimizers excel at single-tenant optimization though kernel-fusion and auto-tuning. However, our approach focuses on optimizing the performance of many disjoint graphs. Our approach is a dynamic alternative to that developed in Guevara et al.~\cite{ETP}, which manually combines small kernels within a stream to increase GPU utilization, yielding up to a 1.3x speedup on a Gaussian Elimination algorithm. While the work in Guevara et al.~\cite{ETP} focuses on manually merging kernels at the CUDA block level, our approach discusses a scalable procedure to batch large numbers of kernels that execute similar matrix multiplication routines together dynamically, as well as interleave CUDA streams. While ours is a matrix-math targeted approach, we demonstrate that it has high scalability on multiple real-world neural network tasks [Table \ref{table:proposed:tput}]. Our approach extends these tools and is complementary to the  existing ecosystem.

\subsection{Benchmarking dynamic space-time kernel scheduling}
\label{subsec:spacetime_superkernel}
We evaluate the overall throughput provided by multiplexing matrix multiplication kernels by time-only, space-only, and a proposed space-time multiplexing strategy. Specifically, we examine the Single Precision floating-point General Matrix Multiply kernel (SGEMM). Matrix multiplication is often used to implement the convolution operator in neural networks, in addition to Fourier-domain, Winograd-domain, and direct kernel implementations \cite{sgemm}.

\begin{figure}[t]
    \begin{minipage}[t]{0.5\textwidth}
        \centering
        \includegraphics[width=\textwidth]{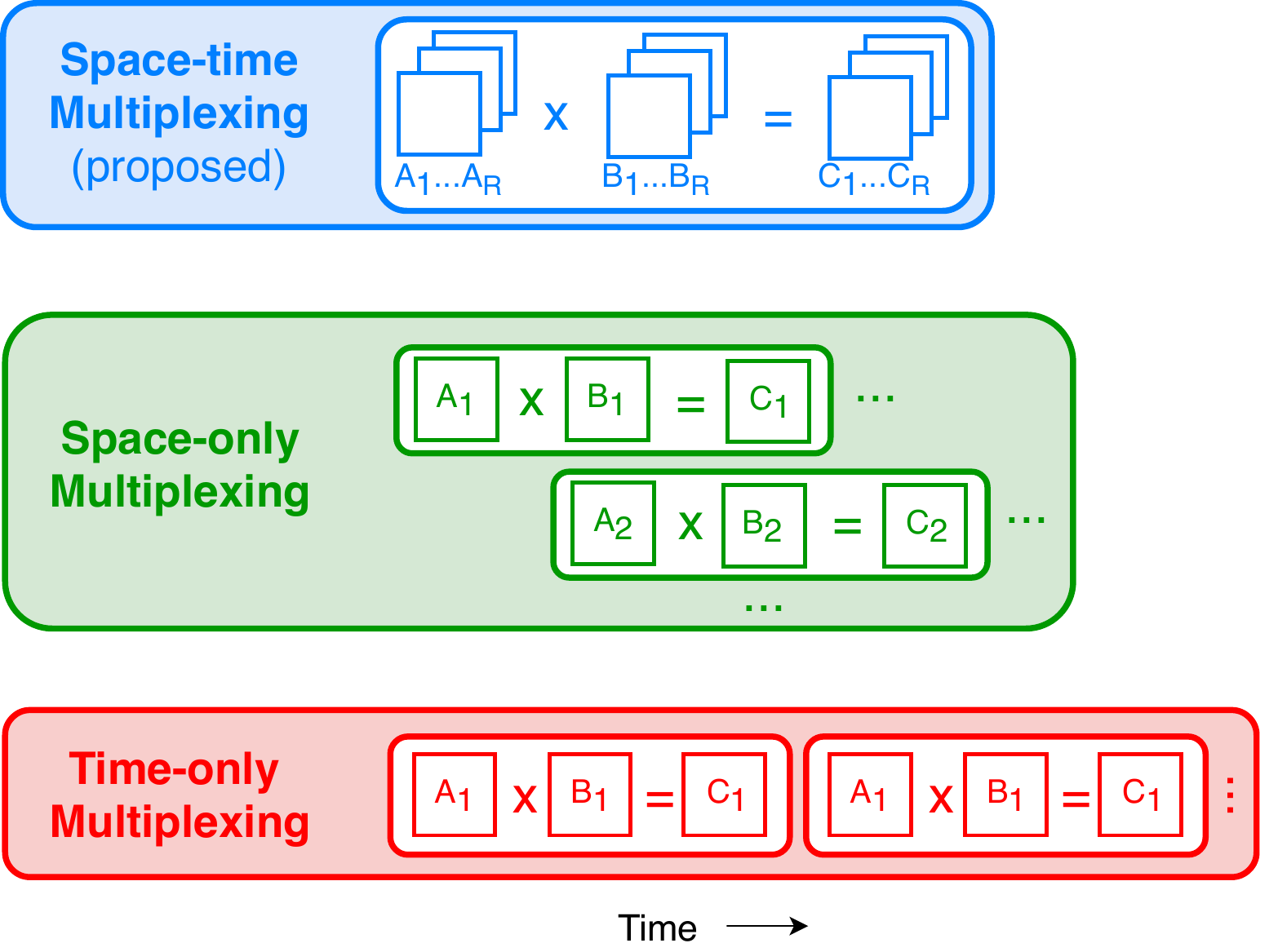}
        \caption{\textbf{Space-time scheduling reduces kernel invocations via inter-model batching.} We illustrate kernel multiplexing methods for $R$ SGEMMs scheduled on the same GPU. Outer boxes depict a single CUDA kernel invocation.}
        \label{fig:kernels}
    \end{minipage}
    \hspace{0.025\textwidth}
    \begin{minipage}[t]{0.48\textwidth}
        \centering
        \includegraphics[width=\textwidth,trim={0cm 0.65cm 0cm 0cm,clip}]{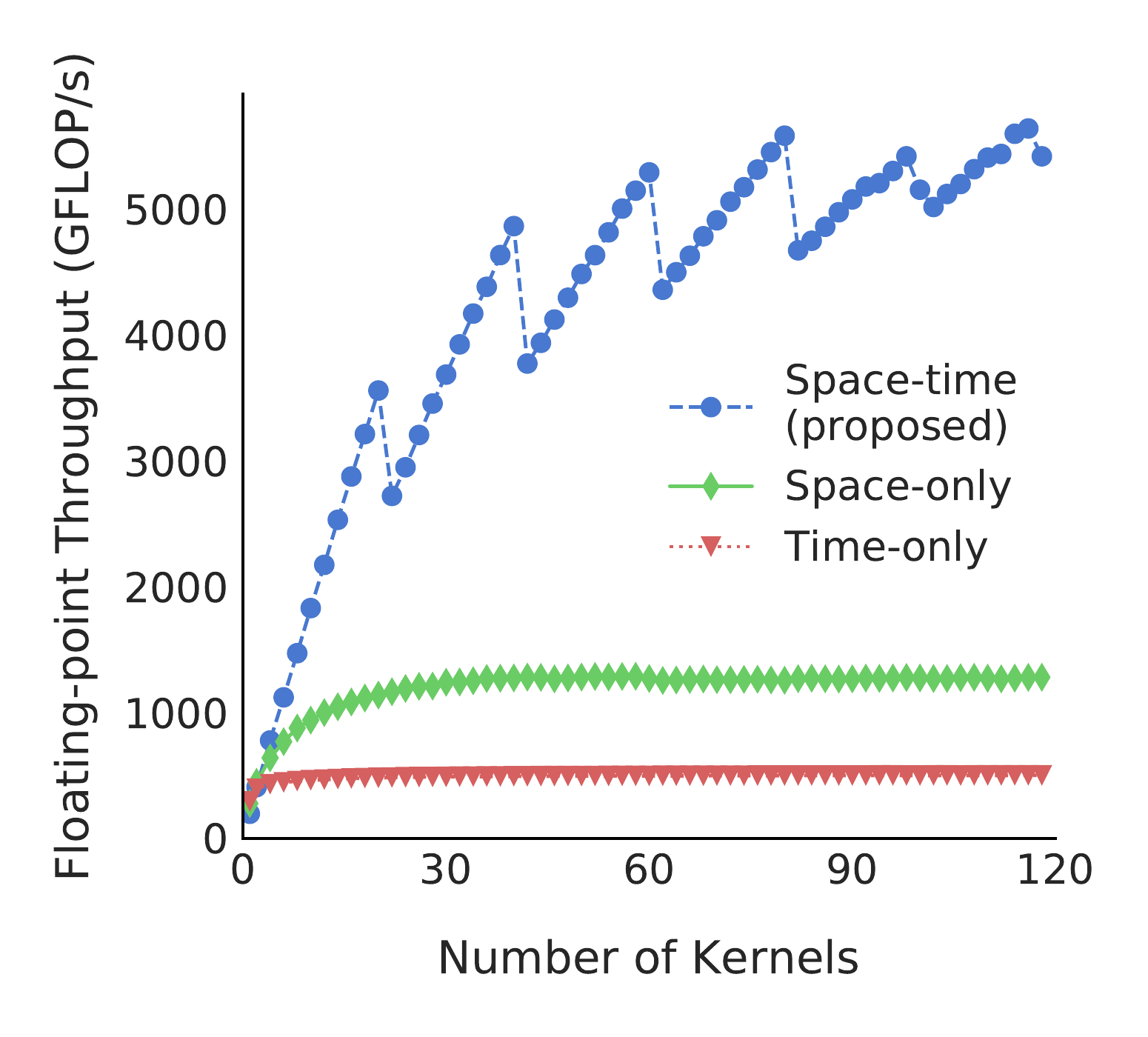}
        \caption{\textbf{Inter-model kernel batching offers ideal convolution throughput scaling.}
        We evaluated the matrix multiply throughputs on an intermediate ResNet-18 convolution kernel.
        % A kernel-level benchmark of matrix multiply throughputs shows the promise of a batched approach. 
        % An intermediate convolution from ResNet-18 is expressed as a SGEMM for this test. 
        % Space-time multiplexing is implemented by evaluating super-kernels with \texttt{cublasSgemmBatched}.
        }
        \label{fig:proposed:sgemm}
    \end{minipage}
\end{figure}

\figref{fig:proposed:sgemm} demonstrates that spatial multiplexing via Hyper-Q/CUDA stream usage can improve throughput as compared to timeslicing approaches to GPU sharing. However, the average throughput is still substantially lower than the single-precision throughput offered by the V100 (\secref{subsec:experimentalsetup}).

%However, there is still a meaningful gap between batching latency and CUDA streams, particularly for small problem sizes. As more tenants share the GPU, this gap grows. Ideally, we would like a solution that scales well as more tenants run on a system.

% \begin{figure}[t]
% %   \begin{minipage}[b]{0.5\textwidth}
% %     \caption{TODO}
% %   \end{minipage}
% %   \begin{minipage}[b]{0.5\textwidth}
%     \centering
%     \includegraphics[width=\textwidth]{figures/sgemm.pdf}
%     \caption{\textbf{Inter-model kernel batching offers ideal SGEMM throughput scaling:} A kernel-level benchmark of matrix multiply throughputs when scheduling increasing numbers of replicas on a single V100 GPU shows the promise of a batched approach. Space-time multiplexing is implemented by evaluating super-kernels with \texttt{cublasSgemmBatched}.
%     %matches single-tenant performance unlike time multiplexing and current spatial multiplexing approaches. Problem sizes are held constant ($M=256,N=256,K=256$), and some overhead is excluded --- these numbers exclude memory-bandwidth effects.
%     \paras{Add superkernel stacking diagram} \ajay{Make square figure, remove Exclusive Access line} \ajay{Remake with other convolution sizes} \todo{RSD: Remove exclusive access + remake w/ other convolution sizes + ResNet18}}
%     \todo{Add explanation why Figure 3 includes exclusive access but Figure 4 does not. Must discuss how it is cost prohibitive, and does not make sense in this context.}
%     \label{fig:proposed:sgemm}
% %   \end{minipage}
% \end{figure}

Instead of maintaining separate kernel streams which the device can schedule at a fine granularity, we investigate a software-based scheduler that batches kernels across models. By batching kernels across many models into a single super-kernel, a single GPU invocation would have an opportunity to saturate all resources on the GPU for its timeslice.

To model the performance of a space-time multiplexing software scheduler, we measure the throughput of a nominal approach --- collecting SGEMM problems from several models of the same architecture into a single batched matrix multiplication super-kernel. These models are have different weights and inputs, as is likely in a multi-tenancy setting. The batched super-kernel is more efficient than many smaller kernel invocations, and also better spatially multiplexes the GPU. This also allows better predictability of latency as the dynamic kernel scheduler can selectively batch kernels and determine when to execute workloads based on per-model SLOs.

\figref{fig:proposed:sgemm} demonstrates substantially better throughput scaling via batching than via time-only and spatial-only multiplexing approaches. Fixing the problem size $M=256,~ N=128,~ K=1152$, the number of concurrently submitted SGEMM problems is scaled. This problem size represents an im2col SGEMM implementation of a representative intermediate convolution in the ResNet-18 architecture (\texttt{conv2\_2}), with a $128 \times 128$ image input to the network, convolutional kernel size $3 \times 3$, and $128$ input and output channels to the layer. Similar floating-point throughput improvements are observed for other intermediate layers, as well as for dense matrix-vector multiplications found in RNNs and square matrix-matrix multiplications (Table \ref{table:proposed:tput}).

\begin{table}[]
\centering
\caption{Space-time scheduling throughput increases over next best approach. $R$ SGEMM kernel evaluations are queued on a NVIDIA V100 GPU.}
\label{table:proposed:tput}
\hspace*{-1.5mm}\begin{tabular}{@{}cccc@{}}
\toprule
 & \begin{tabular}[c]{@{}c@{}}\textbf{Matrix-vector: RNN}\\ \footnotesize{M=512, N=1, K=512}\end{tabular} & \begin{tabular}[c]{@{}c@{}}\textbf{ResNet-18 conv2\_2}\\ \footnotesize{M=256, N=128, K=1152}\end{tabular} & \begin{tabular}[c]{@{}c@{}}\textbf{Square matrix-matrix}\\ \footnotesize{M=N=K=256}\end{tabular} \\ \midrule
$R = 10$ & 1.21x & 1.68x & 2.42x \\ \midrule
$R = 20$ & 2.14x & 2.88x & 2.47x \\ \midrule
\begin{tabular}[c]{@{}c@{}}$2 \leq R \leq 120$ (geomean)\\ \end{tabular} & \textbf{2.48x} & \textbf{3.23x} & \textbf{4.93x} \\ \midrule
\textit{Next best scheduler} & \textit{Time-only} & \textit{Space-only} & \textit{Space-only} \\ \bottomrule
\end{tabular}
\end{table}

This matrix multiply super-kernel is implemented in the NVIDIA cuBLAS operation \texttt{cublasSgemmBatched}. It requires all sub-kernel problem dimensions be the same. However, the MAGMA BLAS library \cite{magma_blas} implements a variable-sized batched SGEMM that would allow for different kernels to be batched. For all compared approaches, data is preallocated on the device as in a real-world DNN inference setting.

\section{Conclusion and Future Directions}
% \todo{Paras rewrite} 
In this work, we evaluated spatial and temporal multiplexing techniques to support inference across multiple models on a single GPU.
We first considered standard approaches utilized by popular DNN frameworks and GPU vendors like NVIDIA.
While these techniques improve utilization, they increase latency and variability in prediction performance in benchmarks. Neither space-only nor time-only mutliplexing techniques could achieve high resource-efficiency, predictable latencies and isolation.

We observe a large performance gap between batch-level parallelism and space-only multiplexing, suggesting substantial opportunities to improve utilization.
We propose a dynamic space-and-time scheduler that addresses all three aforementioned criteria. Software-level fusion of kernel operators across multiple models and inputs presents a promising approach to online inference scheduling.

As an early evaluation of this approach, we studied roof-line performance~\cite{williams2009roofline} available via SGEMM fusion of all queued problems, which offers throughputs that scale well with the number of GPU tenants or model replicas.
We demonstrate a \(>\) 3x speedup compared to the prior state-of-the-art in online inference multitenancy.
We believe this work points towards a new approach to efficient multi-tenant execution of deep neural networks through intelligent inter-model fused kernel scheduling.

\section*{Acknowledgements}
We thank Koushik Sen, Eyal Sela, Zongheng Yang, Anjali Shankar and Daniel Crankshaw for their insightful feedback and edits. In addition to NSF CISE Expeditions Award CCF-1730628, this research is supported by gifts from Alibaba, Amazon Web Services, Ant Financial, Arm, CapitalOne, Ericsson, Facebook, Google, Huawei, Intel, Microsoft, Scotiabank, Splunk and VMware.

\bibliographystyle{plain}
\bibliography{references.bib}

% \section*{References}

% References follow the acknowledgments. Use unnumbered first-level
% heading for the references. Any choice of citation style is acceptable
% as long as you are consistent. It is permissible to reduce the font
% size to \verb+small+ (9 point) when listing the references. {\bf
%   Remember that you can use more than eight pages as long as the
%   additional pages contain \emph{only} cited references.}
% \medskip

% \small

% [1] Alexander, J.A.\ \& Mozer, M.C.\ (1995) Template-based algorithms
% for connectionist rule extraction. In G.\ Tesauro, D.S.\ Touretzky and
% T.K.\ Leen (eds.), {\it Advances in Neural Information Processing
%   Systems 7}, pp.\ 609--616. Cambridge, MA: MIT Press.

% [2] Bower, J.M.\ \& Beeman, D.\ (1995) {\it The Book of GENESIS:
%   Exploring Realistic Neural Models with the GEneral NEural SImulation
%   System.}  New York: TELOS/Springer--Verlag.

% [3] Hasselmo, M.E., Schnell, E.\ \& Barkai, E.\ (1995) Dynamics of
% learning and recall at excitatory recurrent synapses and cholinergic
% modulation in rat hippocampal region CA3. {\it Journal of
%   Neuroscience} {\bf 15}(7):5249-5262.

\end{document}